\documentclass[reprint,
superscriptaddress,
amsmath,amssymb,
 aps,
prd,
floatfix,nofootinbib]{revtex4-2}

\usepackage{geometry}
\usepackage{pstricks-add}
\usepackage{amsmath,amssymb,bbm,bm}
\usepackage{wrapfig}
\usepackage{graphicx}
\usepackage{epstopdf}
\usepackage{color}
\usepackage{hyperref,textcomp}
\usepackage{slashed}
\usepackage{comment}
\usepackage{multirow}

\newcommand{\di}{\text{d}}
\newcommand{\pare}[1]{\left(#1\right)}
\newcommand{\parea}[1]{\left[#1\right]}
\newcommand{\pareb}[1]{\left\{#1\right\}}
\newcommand{\avg}[1]{\left<#1\right>}
\newcommand{\abs}[1]{\left|#1\right|}

\newcommand{\ket}[1]{\left|#1\right>}
\newcommand{\bra}[1]{\left<#1\right|}

\makeatletter
\renewcommand{\lambdabar}{{\mathchoice
  {\smash@bar\textfont\displaystyle{0.25}{1.2}\lambda}
  {\smash@bar\textfont\textstyle{0.25}{1.2}\lambda}
  {\smash@bar\scriptfont\scriptstyle{0.25}{1.2}\lambda}
  {\smash@bar\scriptscriptfont\scriptscriptstyle{0.25}{1.2}\lambda}
}}
\newcommand{\smash@bar}[4]{%
  \smash{\rlap{\raisebox{-#3\fontdimen5#10}{$\m@th#2\mkern#4mu\mathchar'26$}}}%
}
\makeatother

\begin{document}

%\title{Optical tweezer arrays for precision tests of fundamental forces\\Resonant searches for fundamental forces with optical tweezer arrays\\
\title{Probing micron-range fifth forces with cold atoms in optical tweezer arrays}

\author{Marios Galanis}
\email{mgalanis@stanford.edu}
\affiliation{Leinweber Institute for Theoretical Physics at Stanford,
382 Via Pueblo, Stanford, CA, 94305, USA}

\author{Andrew A. Geraci}
\email{andrew.geraci@northwestern.edu}
\affiliation{Center for Fundamental Physics, Department of Physics and Astronomy, Northwestern University, Evanston, IL 60208}
\date{\today}

\begin{abstract}
We propose a new method to search for fifth forces at micron scales, by resonantly exciting the motional state of ground-state-cooled atoms in optical tweezer arrays. We develop a homodyne readout protocol based on sideband thermometry, which can approach the Standard Quantum Limit, and expand the discovery reach for fifth forces by up to four orders of magnitude compared to existing constraints. We show that this level of sensitivity can also probe well-motivated models of extra dimensions beyond the current tabletop state of the art, by leveraging the parametric enhancement of higher-dimensional gravity.

\end{abstract}

\maketitle
\textbf{Introduction---}Submillimeter tests of gravity are a direct discovery channel for physics beyond the Standard Model. Several theories predict that gravitational forces between objects at short distances will differ from the Newtonian inverse-square law due to the presence of new exotic interactions or to the presence of extra spatial dimensions~\cite{add1,dimopoulosguidice,Montero2023}. These theories have motivated a series of experiments and proposals to search for Yukawa-type deviations at laboratory scale distances, involving torsion balances \cite{kapner2007,lee2020}, microcantilevers \cite{Geraci2008}, torsional oscillators \cite{Long2003,Chen2016}, cold atoms \cite{andysavas2003,tinosurface}, levitated superconductors \cite{Paikgravity}, levitated magnets \cite{ulbrichtmaglev}, levitated dielectric particles \cite{Geraci2010,blakemore2021}, neutron interferometry~\cite{PhysRevC.75.015501,doi:10.1126/science.abc2794}, coherent neutron scattering~\cite{PhysRevD.108.055005}, and the M\"{o}ssbauer effect~\cite{PhysRevD.102.115031}.

Atomic systems present several favorable characteristics, such as well-understood electromagnetic interactions with the environment, precise control and manipulation, and absence of solid-state effects that may plague other types of sensors. In this \emph{Letter}, we propose a new method to probe micron-range fifth forces by sensing the motional perturbation of an ensemble of ground-state-cooled atoms in an optical tweezer array due to a moving source mass in close proximity. We develop a readout protocol based on a homodyne variant of sideband thermometry, which almost saturates the quantum Cram\'{e}r-Rao bound and can extend the discovery reach for fifth forces by as much as four orders of magnitude compared to existing constraints. We further show that this level of sensitivity would surpass existing tabletop limits for extra dimensions, due to the parametric enhancement of higher-dimensional gravity. We conclude by discussing major noise sources and commenting on the implementation prospects of our proposal.

%%%%%%%%%%%%%%%%%%%%%
\textbf{Resonant fifth force signal---}Consider an atom of mass $m_a$ in a harmonic trap with angular frequencies $\omega_i=2\pi \nu_i$ and lengths  $R_i\equiv (m_a\omega_i)^{-1/2}$, along the three motional axes\footnote{We use natural units $\hbar=c=1$.}. The atom is prepared in a thermal state $\rho_\text{th}$ with $\bar{n}_i = (e^{\omega_i/T_i}-1)^{-1}\ll1$ thermal quanta, and so occupies predominantly its motional ground state.

\begin{figure}
    \centering
    \includegraphics[width=1\linewidth]{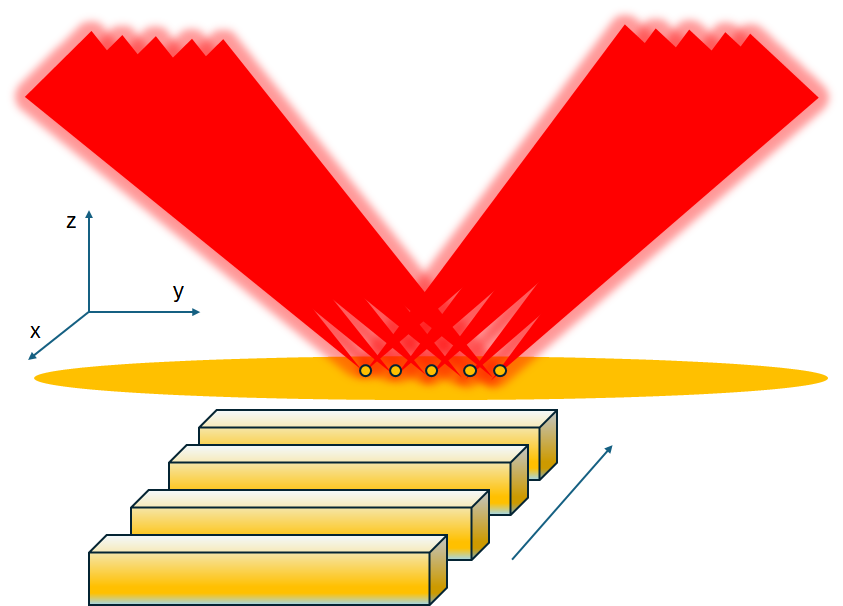}
    \caption{Proposed setup. An array of optical tweezers is used to confine a two-dimensional grid of atoms at submicron range from a thin metallic reflective surface (only a one-dimensional array is shown). Behind the surface, a periodic array of source masses, such as a high-density-contrast sheet of Au and Si, is translated to produce a force on the atoms due to their gravitational interaction plus any fifth force to be searched for.}
    \label{fig:setup}
\end{figure}

The center of the trap is a distance $d$ away from a periodic mass distribution of thickness $h$ made out stripes of width $w$ and density contrast $\Delta \rho = \rho_1-\rho_2$, as depicted in Fig.~\ref{fig:setup}. Using the standard parametrization for a fifth force of strength $\alpha$ and range $\lambda$ of a point mass $M$ compared to Newtonian gravity, $ \Phi = -\frac{G M}{r}\pare{1 + \alpha e^{-r/\lambda}}$, this extended mass distribution sources an $x$-dependent fifth force whose dominant spatial harmonic at the location $x=x_a$ of the atom is

\begin{figure}
    \centering
    \includegraphics[width=1\linewidth]{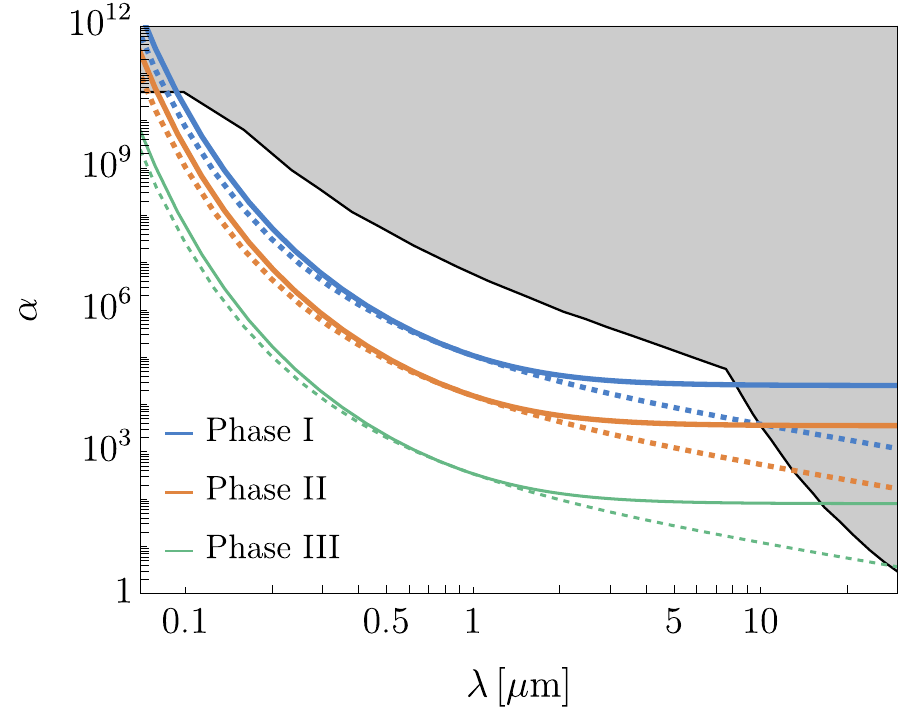}
    \caption{$1~\sigma$ sensitivity of a $^{171}$Yb tweezer array to a resonant Yukawa force of strength $\alpha$ and range $\lambda$, according to Eq.~\eqref{eq:SNR_ideal}. We assume density contrast $\Delta\rho=17~$g/cm$^3$, adequate cooling $2\bar{n}_x+1\approx 1$, $80\%$ readout efficiency, and $T_\text{tot}=100$ days of incoherent averaging. The source-tweezer distance is $d=900~$nm. Parameters for existing technologies (Phase I), near-term capabilities (Phase II), and long-term improvements (Phase III) are given in Table~\ref{tab:phases} below. Optimizing the source periodicity $w$ per $\lambda$ yields the dashed curves, while choosing its optimal value at $1~\mu$m yields the solid curves. The shaded gray region corresponds to leading laboratory constraints~\cite{Chen2016,lee2020}, and neutron star cooling limits~\cite{NScooling}.}
    \label{fig:reach}
\end{figure}

\begin{table}[t!]
  \centering
  \renewcommand{\arraystretch}{1.2} 
  \begin{tabular}{c | r r r | r r r r}
  \hline \hline
 Phase & \multicolumn{1}{c}{$N$} & \multicolumn{1}{c}{$t_\text{shot}$} & \multicolumn{1}{c|}{$\nu_x$} & \multicolumn{1}{c}{$\alpha(1\mu\mathrm{m})$} & \multicolumn{1}{c}{$R_1$} & \multicolumn{1}{c}{$R_2$}  & \multicolumn{1}{c}{$R_3$} \\
\hline
  I   & $10^4$   & $0.1~\mathrm{s}$   & $10~\mathrm{kHz}$ & $1.1\cdot 10^5$   & $4.6~\mathrm{mm}$ & $78~\mu\mathrm{m}$    & $14~\mu\mathrm{m}$  \\
  II  & $10^5$  & $0.5~\mathrm{s}$   & $10~\mathrm{kHz}$ & $1.6\cdot 10^4$   & $0.7~\mathrm{mm}$ & $30~\mu\mathrm{m}$    & $7.6~\mu\mathrm{m}$ \\
  III & $10^6$  & $10~\mathrm{s}$    & $1~\mathrm{kHz}$  & $350$ & $18~\mu\mathrm{m}$ & $5~\mu\mathrm{m}$    & $2.4~\mu\mathrm{m}$  \\
  \hline\hline
  \end{tabular} 
  \caption{Tweezer array parameters for our setup: number of atoms $N$, coherent-integration time $t_\text{shot}$, and radial frequency along the driven axis $\nu_x$. We also give $1~\sigma$ sensitivity benchmarks for the fifth force strength $\alpha$ at one micrometer and the size of $p=1,2,3$ extra dimensions that can be probed, assuming an equal-size toroidal compactification, and the rest of the parameters as in Fig.~\ref{fig:reach}.}
\label{tab:phases}
\end{table}

\begin{equation}
    \Phi_{5,a} = -4\alpha G\Delta \rho \frac{e^{-\kappa_0 d}}{\kappa_0^2}\cos\parea{q_0(x+x_a)}
\end{equation}
where $\kappa_0^2 = q_0^2+\mu^2$, $q_0=\frac{\pi}{w}$, $\mu^{-1}=\lambda$. We have assumed a thick slab, $\kappa_0 h\gg 1$, and tight axial confinement, $\kappa_0 z\ll 1$. Modulating the sheet as $\delta x(t)=x_0 \cos\Omega t$ creates temporal harmonics at $m\Omega$ with strengths $J_m(q_0 x_0)$, where $m$ is an integer. Tuning $\Omega = \omega_x$ resonantly drives the atom.

This is essentially a driven mass on a spring, and so a natural observable to consider is the displacement. Since this is intuitive, we will use it as a proxy for computing the parametric scaling of the SNR in what follows. Our readout protocol, developed later, will not be measuring the displacement directly, but will retain the same scaling. In linear response, the displacement of the atom is given by

\begin{equation}
\delta\!\avg{X_a(t)}
= \alpha  A_G R_x
\mathcal F(T)\,
t\sin\pare{\omega_xt}\cos\pare{\frac{\pi x_a}{w}}
\label{eq:Xdispl}
\end{equation}
where the gravitational-strength amplitude is 

\begin{equation}
    A_G=-4 c_\text{d} Gm_a\Delta\rho\, q_0 R_x\frac{e^{-\kappa_0 d}}{\kappa_0^2}.
    \label{eq:amplitude}
\end{equation}

We took $x_0\sim 1.84 q_0^{-1}$, so that $c_\text{d} = J_1(1.84)\approx 0.58$,
and $\mathcal{F}(T)=
\exp\parea{
-q_0^2R_x^2(2\bar{n}_x+1)/4
}\approx 1$, a thermally-averaged finite-size form factor. The factor $q_0 R_x$ reflects the dipolar nature of the resonant transition. Driving a higher multipole $\ell_x$ would give instead $(q_0 R_x)^{\ell_x}$, which is further suppressed. The amplitude is maximized at $q_0\sim\lambda^{-1}$ for $\lambda\gg d$ and at $q_0\sim (\lambda d)^{-1/2}$ for $\lambda\ll d$. We will assume that $\cos(\pi x_a/w)\simeq 1$ can be engineered by matching the array and source-mass periodicities. 

Given the position variance 
$\text{var}[X_a] = \frac{R_x^2}{2}(2\bar{n}_x+1)$, $N$ independent atoms  coherently driven for $t_\text{shot}$, and $k=T_\text{tot}/t_\text{shot}$ independent repetitions, the signal-to-noise ratio for the displacement at the oscillation maximum is

\begin{equation}
    \text{SNR}=\alpha \eta |A_G|\sqrt{\frac{2 N t_\text{shot}{T_\text{tot}}}{2\bar{n}_x+1}},
    \label{eq:SNR_ideal}
\end{equation}
where we have included a phenomenological readout efficiency factor $\eta$, which we derive for a specific readout protocol later. We plot the fifth force reach in Fig.~\ref{fig:reach}, taking $\eta=80\%$ and $2\bar{n}_x+1\approx 1$ for three tweezer array setups, as summarized in Table~\ref{tab:phases}. We discuss major noise sources and challenges for achieving this sensitivity towards the end of this \emph{Letter}.

As already noted, while displacement is a natural observable, it need not be the only one. An optimal observable extracts the maximum amount of information contained in the quantum state. This is quantified by the quantum Fisher information (QFI), which for $\alpha$ in one shot and $N$ independent atoms is

\begin{equation}
    F_{Q,\alpha}^{(N)}=N\frac{
2A_G^2t_\text{shot}^2}{2\bar{n}_x+1},
\label{eq:QFI}
\end{equation}
assuming $q_0 R_x\ll1$. Indeed, Eq.~\eqref{eq:SNR_ideal} for the displacement saturates the quantum Cram\'{e}r-Rao bound 

\begin{equation}
    \text{var}[\hat{\alpha}]\geq \frac{1}{k F_{Q,\alpha}^{(N)}},
    \label{eq:QCRB}
\end{equation}
for $k$ repetitions, %demonstrating that the motional quadrature constitutes 
and is thus \emph{an} optimal readout of the signal, at $\eta=1$ and at the oscillation maximum. Our measurement protocol will not rely on motional readout, but will similarly be able to almost saturate Eq.~\eqref{eq:QCRB}.

%%%%%%%%%%%%%%%%%%%%%%%

\textbf{Extra dimensions} modify short-range gravity through a Kaluza-Klein (KK) tower of Yukawa forces~\cite{add1}, $\alpha e^{-r/\lambda}\to\frac{4}{3}\sum_{\mathbf n\neq0}e^{-\mu_{\mathbf n}r}$, where $\mu_\mathbf{n}$ are the masses of the KK modes, usually approximated in tabletop experiments by the lightest mode, $\mu_*$, its degeneracy $g$ and $\alpha=\frac{4}{3}g$, in the parametrization of~\cite{pdg,lee2020}. As is well-known, at $d\ll\mu_*^{-1}$ gravity becomes higher dimensional and stronger, $\Phi_G\sim G_{4+p}M/r^{1+p}$, where $G_{4+p}=G\mathrm{Vol}(\mathcal M)$ and $\text{Vol}(\mathcal{M})$ the volume of the $p$-dimensional manifold $\mathcal{M}$. Thus experiments at $d\ll \mu_*^{-1}$ could be sensitive to extra dimensions without gravitational-strength sensitivity.

The sensitivity is given by Eq.~\eqref{eq:SNR_ideal} with $\alpha A_G$ substituted by

\begin{equation}
    \abs{A_{4+p}}=\frac{16}{3} c_\text{d} G\text{Vol}(\mathcal{M})m_a\Delta\rho \,q_0R_xS_p(q_0,d),
    \label{eq:amplitude_extradim}
\end{equation}
%as compared to the Yukawa Eq.~\eqref{eq:amplitude}. 
where $S_p(q_0,d)$ is independent of geometry deep into $\mathcal{M}$. We compute it in App.~\ref{app:extradims}. %The scaling $A_p\propto G_{4+p}$ is the hallmark of the steeper higher-dimensional power-law.

Optimizing numerically for $d=900~$nm, we find $w_1\approx 30~\mu$m, $w_2\approx 6~\mu$m and $w_3\approx 3~\mu$m for $p=1,2,3$ extra dimensions. We take a square torus $T^p$ of radius $R$ as an illustrative example, for which $\text{Vol}(\mathcal{M})=(2\pi R)^p$, and give the $1~\sigma$ radius sensitivity in Table~\ref{tab:phases}. Deep into the manifold, $d,w\ll R_p$, the scaling is $R_p\propto \parea{2Nt_\text{shot}T_\text{tot}R_x^2/(2\bar{n}_x+1)}^{-\frac{1}{2p}}$. In our forecast, we include corrections due to the geometry of the manifold, which are at the tens-of-percent level.
The current constraint from E\"{o}t-Wash is $(R_1,R_2,R_3)\lesssim (30\mu\mathrm{m},26\mu\mathrm{m},24\mu\mathrm{m})$~\cite{pdg,lee2020} for the same compactification, illustrating the potential competitiveness of our proposal. Importantly, this signature applies also to other setups which forecast similar sensitivities at micron-scale separations, such as optomechanical sensors~\cite{Geraci2010, blakemore2021,Dana:2026}. 

Astrophysical and collider searches exploit the same KK multiplicity through \emph{on-shell} production up to MeV and TeV energies, respectively. This is many orders of magnitude away from the scale of interest, which we can probe more directly. In the ADD model with a square $T^p$, colliders give the stronger bounds $(R_2,R_3)\lesssim(3.8\mu{\rm m},48{\rm pm})$~\cite{pdg}. Astrophysical and cosmological constraints can be substantially stronger, albeit sometimes model-dependent~\cite{pdg}.

\textbf{Experimental Setup---}We consider a setup with an array of optical tweezers derived from a common high-power laser source that is split into a two dimensional grid of trapping sites with a $9~\mu$m nearest-neighbor lattice spacing. Tweezers of several thousand atoms have been demonstrated; for instance Refs.~\cite{Manetsch_2025,Wang_2025} report 6100 and 11,000 trapped atoms, respectively. For concreteness we choose $^{171}\text{Yb}$ atoms, although analogous setups may be achievable with other atomic species such as Sr. As a possible method to position a two-dimensional atomic array in close proximity to a surface, 
an array of atoms could be loaded into the two-dimensional grid and the lasers reflected at 45 degrees from a gold-coated silicon nitride (SiN) membrane which acts as a mirror, as shown in Fig. \ref{fig:setup}. By translating the membrane surface towards the laser array, the atoms can be confined in close proximity to the surface.  For a linearly polarized laser with a $p$-polarization configuration, minimal interference will occur between the incident and reflected beam, and the atomic array can be positioned at an arbitrary distance from the reflecting mirror. By rotating the polarization of the laser \textit{in situ}, a standing wave can form in the vicinity of the surface leading to tighter confinement and higher trapping frequencies along the axial direction. A similar approach was recently studied to trap nanoparticles in an optical tweezer near a surface using a 45-degree tilted mirror trap \cite{Grinin2026}.

We consider transverse atom trap frequencies in the range $1$--$10$~kHz, corresponding to trap sizes (i.e. ground-state oscillator lengths) of $R_x\simeq 0.24$--$0.075~\mu$m, respectively. Larger trap sizes enhance the signal due to larger dipoles, but at the same time the linewidth requirements on the lasers and the cooling of the atoms become more stringent. This frequency range remains within current experimental capabilities \cite{PhysRevX.12.021027}, and for example could be realized with a tweezer array using a laser with wavelength of $\lambda=2\pi/k_L=532$ nm, waist size $w_0=
1.0~\mu$m and power $P=0.7 ~m$W  and peak intensity $I_0=2P/(\pi w_0^2)$, leading to a trap depth 
$U_0 = -\frac{1}{4}\alpha_g I_0$
 where $\alpha_g = 37.9$ Hz/(W cm$^{-2}$) is the ground state polarizability at 532 nm \cite{Yamamoto_2016}. This results in frequencies in a standing wave trap of $\omega_{x,y}=\sqrt{4U_0/mw_0^
2} = 2 \pi \times 10$ kHz and $\omega_z=\sqrt{2U_0k_L
^2/m} = 2 \pi \times 84$ kHz.

%%%%%%%%%%%%%%%%%%%%%
\textbf{Homodyne sideband thermometry (HST) readout---}
Reading out a motional quadrature requires a cavity, which complicates the engineering of our proposal and can introduce significant technical noise. Instead, we propose to interfere the fifth-force signal with a pulse of known phase and amplitude, and use sideband thermometry, a well-established technique in optical tweezer arrays, to read out the occupation number of the excited motional state. Related ideas have been demonstrated in trapped-ion molecular spectroscopy ~\cite{Sinhal_2020,najafian20,Meir_2019}.

In standard sideband thermometry, the atoms are initialized in the ground state of an internal degree of freedom with splitting $\omega_{ge}$. Raman pulses at $\omega_\text{B,R} = \omega_\text{ge}\pm \omega_x$ excite the transitions $\ket{g,n_x}\to\ket{e,n_x\pm 1}$, called the blue and red sidebands, respectively. Importantly, the ground state $\ket{g,0}$ is dark to the red sideband. The probabilities to be in $\ket{e}$, for weak pulses, are $p_\text{R}\approx C\avg{n}$ and $p_\text{B}\approx C(\avg{n}+1)$, where $C$ is related to the area of the pulse. These can be measured with state-dependent fluorescence, for instance, and follow binomial distributions $\text{var}[\hat{p}_i] = p_i(1-p_i)/N_i$ for $i=$~R, B. The sideband asymmetry $\mathcal{R}\equiv\frac{p_\text{R}}{p_\text{B}-p_\text{R}}=\avg{n}$ then measures the population of the excited state. For a weak signal of area $\gamma$ the excitation probability is quadratic in it, $\avg{n}=\bar{n}_x+\gamma^2$. For the Yukawa signal, for instance, $\gamma=\alpha A_G t_\text{shot}/\sqrt{2}$. The ratio of the Fisher matrix for this measurement compared to the QFI is $\frac{F_{\alpha\alpha}}{F_{Q,\alpha}}\approx\frac{\gamma^2}{\bar{n}_x}$, which would require unattainable cooling to saturate the  quantum Cram\'{e}r-Rao bound, Eq.~\eqref{eq:QCRB}.

Our \emph{homodyne} protocol consists of applying a coherent displacement $s\beta$ to the atoms, parallel  ($s=+1$) and antiparallel ($s=-1$) to the fifth-force signal, whose phase is known. This can be achieved with negligible heating using two far-detuned beams to create an optical dipole force at $\omega_x$, or by shaking the tweezer centers at $\omega_x$~\cite{McCormick_2019}. In this case $p_i^\pm\approx C(\bar{n}_x+\beta^2+2s\gamma\beta+\gamma^2+\delta_{\text{B},i})$. In the ensemble of $N$ atoms, we assign $N_\text{B}^+$, $N_\text{B}^-$ and $N_\text{R}^+$ atoms to the three sidebands, and define the observable $\mathcal{R}_\text{HST}\equiv \frac{p_\text{B}^+-p_\text{B}^-}{p_\text{B}^+-p_\text{R}^+}\approx4\beta\gamma$, where both $\bar{n}_x$ and $C$ drop out, to the degree that they are common-mode. This is \emph{linear} in $\gamma$ and independent of $\bar{n}_x$. The Fisher matrix is $F_\gamma^{(N)} \approx 4C\beta^2N_\text{B}$ for $\bar{n}_x\ll\beta^2\ll 1$ and $C\ll 1$,
where $N_\text{B}=N_\text{B}^++N_\text{B}^-$. For small $\gamma$, the red-sideband atoms only calibrate $(\bar{n}_x,C)$ and so we may assign $N_\text{R}\ll N_\text{B}\approx N$. 

\begin{figure}
    \centering
    \includegraphics[width=1\linewidth]{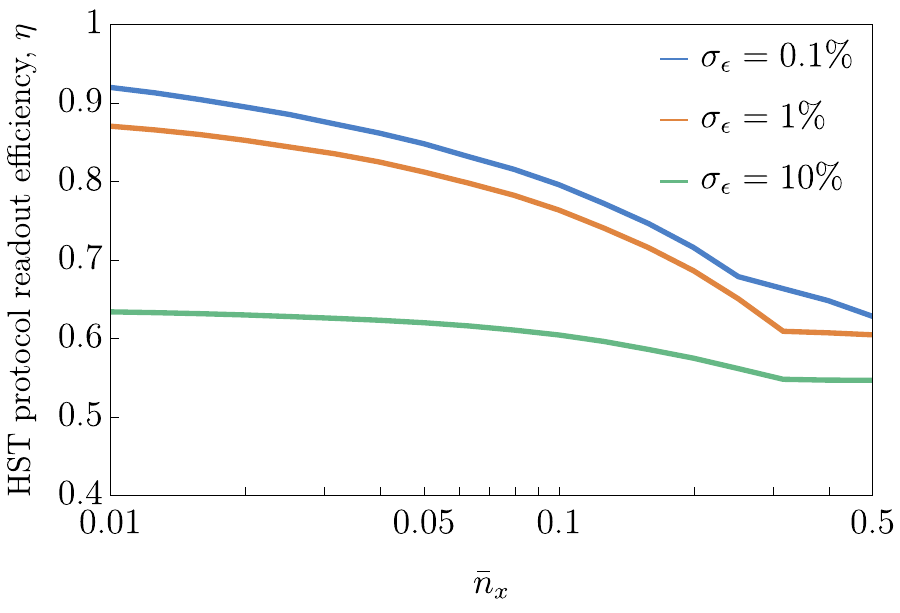}
    \caption{Square-root of the ratio of the marginalized Fisher matrix under the homodyne sideband thermometry readout protocol with respect to the QFI, optimized against the homodyne and Raman pulses, as discussed in App.~\ref{app:homodyne}.
    The vertical axis is equal to the readout efficiency $\eta$ in Eq.~\eqref{eq:SNR_ideal}.
    We have included a Raman pulse area error $\sigma_\epsilon$ to suppress artificial large-area pulses. The breaking features at large $\bar{n}_x$ reflect discontinuities in the numerical optimization.  }
    \label{fig:FisherRatio}
\end{figure}

In practice, we need finite pulses with $C,\beta\sim \mathcal{O}(1)$ to saturate the quantum Cram\'{e}r-Rao bound. This calculation is less transparent, and is presented in App.~\ref{app:homodyne}. In Fig.~\ref{fig:FisherRatio} we plot the ratio of the Fisher matrix of the HST protocol to the QFI, after marginalizing over finite $(C,\bar{n}_x)$, and optimizing for finite $\beta$. To achieve the $80\%$ readout efficiency $\eta$ assumed in Fig.~\ref{fig:reach}, heating should be curtailed to $\bar{n}_x\lesssim 0.1$. The HST protocol trades the extreme cooling requirements on $\bar{n}_x$ for the need to calibrate uncorrelated homodyne pulse errors $\text{var}[\delta\beta_a]\lesssim 1$ for atom $a$, and common errors between all atoms to $\text{var}[\delta\beta_\text{com}]\lesssim N^{-1}$, per shot.

%%%%%%%%%%%%%%%%%%%%%
\textbf{Noise---}
In the remainder, we estimate tolerances for the dominant backgrounds and argue that Phase I may be already feasible with existing technologies. We conclude by commenting on the near- and far-term prospects for Phases II and III.

Let $V_N$ be the potential of a noise source, such that
$    V_N(x_a+X,t)
\approx
\left.(V_N
+
X\,\partial_xV_N+
\frac{1}{2}X^2\,\partial_{x}^2 V_N)\right|_{x=x_a}.$ The second term is a forcing $F(t)=-\partial_x V_N$, and the third an angular frequency correction $\Delta(t) \approx \frac{\partial_x^2 V_N}{2m\omega_x}$. A non-zero mean, $\bar{\delta}=2\pi\delta\nu=\avg{\Delta(t)}$, is a static detuning. Subtracting the means and taking the one-sided power spectra $S(\omega)$ we derive a Lindblad equation for the atom in App.~\ref{app:lindblad}. Heating obeys

\begin{equation}
    \frac{\di \bar{n}_x}{\di t} = \frac{S_F(\omega_x)}{4m\omega_x} + S_\Delta(2\omega_x)\pare{\bar{n}_x+\frac{1}{2}},
    \label{eq:heating}
\end{equation}
corresponding to resonant and parametric heating, respectively. The latter involves an additional derivative, so tolerances on $S_F(\omega)$ should be understood as imposing similar ones on $S_\Delta(2\omega_x)$.

Frequency shifts impede coherent accumulation of the signal due to dephasing, suppressing its amplitude $|\gamma|$ by a factor

\begin{equation}
\epsilon_\gamma
=
\left|
\frac{
1-e^{-\parea{S_\Delta(0)/2+i\bar\delta}t_{\rm shot}}
}{
\parea{S_\Delta(0)/2+i\bar\delta}t_{\rm shot}
}
\right|.
\label{eq:dephasing}
\end{equation}

In general dephasing depends on the full power-spectrum over the integration window (App.~\ref{app:lindblad}), but we assume it to be white here for our estimates.

To derive tolerances we use Phase I tweezer parameters and impose $\bar{n}_x\lesssim 0.1$ or heating rate $\Gamma_x\lesssim 1$~phonon/s, as well as $\epsilon\approx 1$. This  would allow for $\eta\approx80\%$, taking the blue or orange lines of Fig.~\ref{fig:reach}.  We emphasize that the heating requirements on the orthogonal directions to the signal, $y$ and $z$, can be much less stringent.

The atoms couple to dominant backgrounds electromagnetically, through their polarizability $\alpha_0$ and magnetic moment $\mu$. For $^{171}$Yb we take $\alpha_0\approx 21~\text{\AA}^3$ and $\mu\approx 0.5\mu_N$, where $\mu_N$ is the nuclear magneton. The atom-shield separation is set to $z_s=500$~nm. There are two electrostatic potentials between the atom and the shield, Casimir-Polder and patch potentials,

\begin{equation}
\begin{split}
    V_\text{CP}(z_s)&=-\frac{3}{8\pi}\frac{\alpha_0}{z_s^4},\qquad V_\text{p} = -2\pi \alpha_0\avg{E^2}.
\end{split}
\end{equation}

The Casimir-Polder potential arises from the vacuum fluctuations of the electromagnetic field between the atom and the shield. The patch potential arises from surface imperfections of the shield that lead to spatially varying electrostatic potentials. The spatial correlations of the patches are hard to model and, in practice, have to be measured for a given shield. As a conservative estimate, we will take $E_\text{p}\sim\phi_\text{rms}/z_s$, with size $\phi_\text{rms}\sim 10$~mV for gold surfaces~\cite{Sushkov2011, Garrett_2015}. If instead the typical patch size is $\ell\ll z_s$, then local electric fields are further suppressed by $\ell/z_s$. For the parameters considered here, $V_\text{CP}\sim 8~$peV and $V_\text{p}\sim 3$~peV.

\emph{Shield motion---}If there is any fluctuation $z_s\to z_s+w(\mathbf{r},t)$, both CP and patch potentials contribute primarily forcing noise. 
A stochastic tilt $w=\theta(t)x$ requires $\sqrt{S_\theta(\omega_x)}\lesssim 2\times 10^{-4}~\text{rad}/\sqrt{\text{Hz}}$. 
The tweezer beams are reflected off the shield 
and a piston fluctuation $z\to z+\delta z$  
can  heat the $z$-motion by changing the interference pattern of the incident and reflected beams. While orthogonal to the signal direction, $xz$ interactions, such as 
$\delta V=m\omega_z^2 xz\delta\theta$, can move thermal quanta to $x$. Taking $\bar{n}_z\lesssim 1$, we get $\sqrt{S_{\delta z}(\omega_z)}\lesssim 9~\text{pm}/\sqrt{\text{Hz}}$ at $\nu_z=10~$kHz or $\omega_z^2\sqrt{S_{\delta z}(\omega_z)}\lesssim 2\times 10^{-2}~\text{m/s}^2/\sqrt{\text{Hz}}$ \emph{relative} acceleration noise. The shield should have rigid mounts, and be made out of a high-tension membrane, such as SiN, to raise its fundamental frequency and suppress any response at $\omega_x$.

\emph{Time-dependent patch potentials---}The patch potentials may exhibit temporal fluctuations due to adsorbate dynamics and other fluctuating surface processes, leading to a forcing $F=4\pi\alpha_0E_\text{p}\partial_x\delta E_\text{p}$. The requirement is $\sqrt{S_{\partial_x\delta E_p}(\omega_x)}\lesssim 70~\text{MV/m}^2/\sqrt{\text{Hz}}$ or $\sqrt{S_{\delta E_p}(\omega_x)}\lesssim 22~\text{V/m}/\sqrt{\text{Hz}}$, assuming $\partial_x\sim z_s^{-1}$. Recent measurements~\cite{Dana:2026} with a charged levitated nanosphere directly constrain surface electric-field noise at $\sim~$kHz frequencies and $z_s\sim 1~\mu$m separations. Their noise floor is consistent with thermal force noise, which allows them to place an \emph{upper bound} $\sqrt{S_{\delta E_p}}\lesssim0.8~{\rm V/m}/\sqrt{\rm Hz}$. Even at $\nu_x=1$~kHz and $t_\text{shot}\sim 10~$s, this bound can satisfy our requirements. 

Slow drifts of the patches may induce dephasing due to their support at low frequencies $\nu_\text{min}\sim1/t_\text{shot}= 10$~Hz. The $\epsilon_\gamma\sim 1$ requirement yields $\sqrt{S_{\delta E_\text{p}}(\omega_\text{min})}\lesssim 90$~V/m/$\sqrt{\text{Hz}}$. To our knowledge, no experiment has measured this noise in the relevant regime. 

All patch potential estimates involve high derivatives of the spatiotemporal patch potential power spectrum, which is highly uncertain. This noise should ultimately be characterized by experiment.

\emph{Johnson noise---}The finite shield temperature gives rise to magnetic noise with power spectrum $S_{B_\text{J}}(\omega)=\frac{T_\text{s}}{16\pi\varrho_\text{Au} z_s}\frac{h_\text{c}}{z_s+h_c}$~\cite{JohnsonNoise,MagneticJohnson},
where $\varrho_\text{Au}$ is the resistivity of gold and $h_\text{c}$ the shield thickness. Heating depends on $S_{\partial_x B_\text{J}}(\omega)\sim z_s^{-2} S_{B_\text{J}}$, where we took $\partial_x\sim z_s^{-1}$. At $20^\text{o}$~C and $h_\text{c}=200$~nm, the heating of $^{171}$Yb is $\Gamma_{B_\text{J}}\sim 10^{-7}~\text{s}^{-1}$. This is sufficiently long, but precludes the use of atoms with electronic spin for Phases II and III, for which $\Gamma_{B_\text{J}}^{(e)}\sim 1~\text{s}^{-1}$.

\emph{Laser noise} contributes to heating and can be estimated from Ref.~\cite{Thomasheating}. Intensity fluctuations induce parametric heating with rate $\Gamma_I=\frac{\omega_x^2}{4} S_I(2\omega_x)$, while pointing fluctuations induce forcing noise, $\Gamma_x=\frac{m_a\omega_x^3}{16\pi^2}S_x(\omega_x)$, from which we can estimate the requirements as $\sqrt{S_I(2\omega_x)}\lesssim 2\times 10^{-5}/\sqrt{\rm Hz}$ or $\text{RIN}(2\nu_x)\lesssim -93~$dBc/Hz, and $\sqrt{S_x(\omega_x)}\lesssim 13~{\rm pm}/\sqrt{\rm Hz}$. At $1~$kHz, these are instead $-73$~dBc/Hz and 490~pm/$\sqrt{\text{Hz}}$. Ref.~\cite{Wang:20} has demonstrated RIN at the level of $-120$~dBc/Hz at frequencies 1--20~kHz, corresponding to heating rates of $10^{-3}$~phonons/s at 10~kHz, comfortably below our requirements.

\textit{Recoil heating---}Optical pumping during sideband cooling produces recoil heating, leaving a residual $\bar{n}\approx \frac{N_\text{sc}}{3}\frac{k^2}{2m_a\omega_x}$ for $N_\text{sc}$ scattered photons of wavelength $\lambda=\frac{2\pi}{k}$ and isotropic emission, with the pump perpendicular to the measuring axis. For $^{171}$Yb, taking $N_\text{sc}\approx 3$ and $\lambda=556$~nm, gives $\bar{n}\approx 0.4 (4)$ at 10(1)~kHz. Direct preparation of the trap could be challenging for softer traps, so a potential protocol would cool the atoms at a tighter trap, and adiabatically expand it on a timescale $\tau_\text{ad}$, such that $\omega_x^{-1}\ll \tau_\text{ad}\ll t_\text{shot}$, to the desired frequency. This preserves $\bar{n}$ in principle. Ref.~\cite{PhysRevX.12.021027} has achieved $\bar{n}\approx 0.08-0.15$ at 27--140~kHz for 100-atom $^{171}$Yb array. We leave considerations of technical noise for such a scheme to future work.

In addition, residual amplified spontaneous emission (ASE) near atomic resonances can cause recoil heating and should be characterized experimentally. This contribution can be mitigated by spectral filtering, for example with an etalon or a suitable atomic vapor cell that absorbs the near-resonant ASE, as in Ref~\cite{ASEloss}. Other methods of filtering ASE have also been employed, which make use of single mode optical fiber \cite{Voigt2001}.

\emph{Dephasing---}Coherent accumulation over 0.1~s requires static frequency shifts $\delta\nu\lesssim 1.6~$Hz, fractionally $1.6\times 10^{-4}$ at 10~kHz and drifts at $\sqrt{S_\Delta(\omega_\text{min})}\lesssim 2.2~$Hz/$\sqrt{\text{Hz}}$. The  translate into requirements for the laser's power and waist inhomogeneities, $\pare{\frac{\delta P}{P},\frac{\delta w}{w}}\lesssim \pare{3,0.8}\times 10^{-4}$, and $\text{RIN}(\nu_\text{min})\lesssim -67~\mathrm{dBc}$/Hz.  

These tolerances apply \emph{only} along the driven direction, and so one scalar depth control per tweezer should be sufficient for static frequency tuning. Ref.~\cite{Chew_2024} has achieved $\delta\nu\lesssim 10~$Hz motional frequency homogenization of a 72-tweezer $^{87}$Rb array, and the measurement was limited by their frequency resolution. Frequency uniformity is not a fundamental requirement, since the atoms are independently interrogated during readout, and so alternative excitation schemes could be developed, which may relax this requirement. 

\emph{Background gas collisions} eject atoms from their tweezers, reducing their effective number to $N_{\rm eff}\simeq N\exp(-t_{\rm shot}/\tau_{\rm vac})$. Requiring $N_{\rm eff}\gtrsim0.9N$ requires a vacuum-limited lifetime $\tau_{\rm vac}\gtrsim1$~s, or pressures $P\lesssim5\times 10^{-9}$~mbar, routinely achievable in ultrahigh-vacuum cold-atom experiments.

\textbf{Systematics---}Electromagnetic fields can leak through the shield and exert a resonant and coherent forcing on the atoms. This leakage needs to be strictly smaller than the fifth force signal.

\emph{Finite conductivity---}The finite conductivity of the shield allows electric fields sourced by the oscillating mass to leak through. We take a fixed potential difference $\Delta\phi=100~$mV in the Au--Si interface. The electric field at the position of the atom is screened by the resistivity $\varrho$ in the thin-shield limit $q_0 h_\text{s}\ll 1$, given by $E_a=\frac{4c_\text{d}\Delta\phi}{\pi}\frac{\varrho \omega_x e^{-q_0 z_s}}{qh_\text{s}\sinh(q_0a)}$ with $a=d-h_\text{s}-z_s$ the source-shield vacuum gap~\cite{Melcher1981}.
The systematic driving potential is $\delta V_E \sim -4\pi\alpha_0 E_{\rm DC}E_{\rm source}\mathcal{S}_E\approx 10^{-23}~$eV, for $h_\text{s}=200$~nm. In contrast, the fifth force signal at a micron has $V_\text{Y}\sim 4\alpha\cdot 10^{-24}$~eV and our sensitivity is at $\alpha\gg 1$.

\emph{Magnetic susceptibility---}Stray magnetic fields polarize the Au--Si segments of the source mass, giving rise to a patterned magnetic field due to their differential magnetic susceptibility, $\Delta\chi\approx 3\cdot 10^{-5}$. 
An unshielded $B_0$ induces a coherent resonant potential of size $V_B\approx\mu \frac{B_0 \Delta\chi}{\pi}e^{-q_0 d} \cos\parea{q_0(x+x_a)}$. For this to be subdominant to the fifth-force signal, we need $B_0\lesssim 0.1~\mathrm{G}\pare{\frac{\alpha}{10^5}}$, which requires magnetic shielding, especially for Phases II and III, or using atoms with vanishing linear magnetic moment, such as $^{174}$Yb or $^{88}$Sr. Alternatively, by considering tailored alloys of material, it may be possible to suppress $\Delta\chi$ by an order of magnitude or more. A similar approach was taken for the test masses used in LISA Pathfinder by using an alloy of gold and platinum \cite{lisamagnetic}.

\emph{Finite shielding---}The Au--Si material contrast produces a modulated CP potential, which can leak through the shield. The CP interaction is not screened by the conductivity of the shield, but by its response at high imaginary frequencies. The finite-thickness Lifshitz calculation of Ref.~\cite{andysavas2003} found that a thickness $h_{\rm c}\gtrsim 200~\mathrm{nm}$ suppresses this potential by at least $10^{-9}$, and thus below our sensitivity. A modestly thicker shield, or an additional Au coating on the segmented source would also provide a substantial safety margin.

\emph{Prospects---}From the above discussion, it appears that the principal ingredients for Phase I have been demonstrated separately. If successfully combined, 10 days of averaging could already explore up to an order of magnitude of open parameter space. Once the somewhat stringent frequency uniformity requirements have been achieved, scaling arrays up to Phase II parameters should be feasible near term, as, for instance, Ref.~\cite{Holman600} has already demonstrated a $600\times 600$, unloaded, optical array. Achieving the ultimate sensitivity of Phase III requires significant advances in several directions, such as much more stringent frequency uniformity conditions, and cooling and stability schemes for softer traps. The scaling-up of tweezer number could be achieved with independent modules, rather than a single extended array under very high laser power. 

\textbf{Discussion---}In this \emph{Letter} we proposed a new method to search for fifth forces by resonantly exciting the motional state of cold atoms in optical tweezer arrays and reading them out with a homodyne variant of sideband thermometry. Our proposal may extend the discovery potential of new micron-range forces by several orders of magnitude, and ultimately probe models of extra dimensions beyond the current state of the art of tabletop experiments.  

A tweezer-array-based sensor directly benefits from the rapid progress in optical tweezers for quantum computation. Continued advances, including integration with microcavity arrays~\cite{tweezercavity, Ding:26}, scaling up of tweezer arrays, taming of technical noise, and quantum-enhanced sensing protocols, could improve readout efficiency and help extend the fifth-force reach closer to gravitational strength at a micron.

\textbf{Acknowledgements---}We are very grateful to Junwu Huang, Tim Kovachy and Ken Van Tilburg for comments on the draft. We also thank Asimina Arvanitaki, Savas Dimopoulos, Peter Graham, Michael Fedderke, Junwu Huang and Ken Van Tilburg for useful discussions. MG acknowledges support from NSF Grant No. 2310429 and
Simons Foundation Grant (824870, PG). A.G.~acknowledges support from NSF grants PHY-2409472 and PHY-2111544, DARPA, the John Templeton Foundation, the W.M.~Keck Foundation, the Gordon and Betty Moore Foundation Grant GBMF12328, DOI 10.37807/GBMF12328, the Alfred P.~Sloan Foundation under Grant No.~G-2023-21130, and the Simons Foundation.
%%%%%%%%%%%%%%%%%%%
\appendix

\section{Finite-pulse homodyne sideband thermometry protocol}
\label{app:homodyne}

In the Lamb-Dicke regime $q_0R_x\ll 1$, the atom experiences a linear forcing from the fifth force, $V(t) =x F_5\cos\omega_x t$, as can easily be seen, for instance, from Eq.~\eqref{eq:phiextradim}. A homodyne pulse, also in the Lamb-Dicke regime, is similarly linear, $V_\text{h} = x\,s F_\text{h} \cos\omega_x t$, $s=\pm 1$. Thus the time evolution under the subsequent application of the two pulses becomes a total coherent displacement,

\begin{equation}
    D(\eta) = e^{\eta a^\dagger - \eta^* a},\quad \eta\equiv -i \frac{R_x}{2\sqrt{2}}(sF_\text{h}t_\text{h} + F_5 t_\text{shot}),
\end{equation}
where the time $t_\text{h}$ that the homodyne pulse is applied for is short compared to $t_\text{shot}$, and $s=\pm$. We define the pulse areas $\beta\equiv \frac{R_x}{2\sqrt{2}}F_\text{h}t_\text{h}$ and $\gamma\equiv \frac{R_x}{2\sqrt{2}}F_5t_\text{shot}$ as in the main text. The thermally displaced motional part of the density matrix is

\begin{equation}
    \rho_\text{TD} = D(\eta)\rho_\text{th}D^\dagger(\eta) = \sum_{n,m} p_{nm}\ket{n}\bra{m},
\end{equation}
where $p_{nm} = \sum_k p_k^\text{th}\bra{n}D(\eta)\ket{k}\bra{k}D^\dagger(\eta)\ket{m}$, and $p_k^\text{th}=(1-e^{-\omega/T})e^{-k\omega/T}$, the thermal probabilities. Including the internal state, the full density matrix is $\rho = \rho_\text{TD}\otimes\ket{g}\bra{g}$.

The blue and red thermometry pulses are modeled as $V_\text{R}(t)=\frac{\Omega(t)}{2}(e^{i\phi}a\sigma_+ + e^{-i\phi}a^\dagger \sigma_-)$ and $V_\text{B}(t)=\frac{\Omega(t)}{2}(e^{i\phi}a\sigma_- + e^{-i\phi}a^\dagger \sigma_+)$, where $\sigma_+=\sigma^\dagger_-=\ket{e}\bra{g}$. These Raman pulses mix only $\ket{g,n}$ and $\ket{e,n\pm 1}$ for each $n$, so that

\begin{equation}
\begin{split}
    U_{\text{B}}\ket{g,n}= &\,\cos\frac{\theta\sqrt{n+1}}{2}\ket{g,n}\\
    &\,- i e^{-i\phi}\sin\frac{\theta\sqrt{n+1}}{2}\ket{e,n+1}\\
    U_{\text{R}}\ket{g,n}= &\,\cos\frac{\theta\sqrt{n}}{2}\ket{g,n}- i e^{-i\phi}\sin\frac{\theta\sqrt{n}}{2}\ket{e,n-1},
\end{split}
\end{equation}
where $\theta = \int\Omega(t')\di t'$ is the area of the Raman pulse on resonance, and $U_\text{R}\ket{g,0}=\ket{g,0}$. The final density matrix is $\rho_\text{B,R}=U_\text{B,R}\rho_\text{TD}\otimes\ket{g}\bra{g}U^\dagger_\text{B,R}$, and the sideband probabilities are the projection into the excited internal state, $p_{B,R}=\text{Tr}[\ket{e}\bra{e} \rho_\text{B,R}]$, which yields

\begin{equation}
    p_i\equiv \sum_{n=0}^\infty P_n\sin^2\pare{\frac{\theta\sqrt{n+\delta_{\text{B},i}}}{2}},
\end{equation}
with $P_n\equiv \bra{n}D(\eta)\rho_\text{th}D^\dagger(\eta)\ket{n}$. This is given by a standard result

\begin{equation}
    P_n(x,\bar{n})=\frac{1}{\bar{n}}\pare{\frac{\bar{n}}{1+\bar{n}}}^{n+1} e^{-\frac{x}{1+\bar{n}}}L_n\parea{-\frac{x}{\bar{n}(1+\bar{n})}}
\end{equation}
where $L_n$ are Laguerre polynomials and $x=(s\beta + \gamma)^2$.

The set of parameters we want to infer is $\boldsymbol{\vartheta}=(\gamma,\theta,\bar{n}_x)$. We further define the subspace $\boldsymbol{\nu}=(\theta,\bar{n}_x)$ to be marginalized over. Here we neglect sources of noise entering through $\beta$. The HST observables are $\mathbf{p}=(p_\text{B}^+,p_\text{B}^-,p_\text{R}^+)$. At vanishing $\gamma$, we see that $\left.\partial_\gamma p_\text{B}^\pm\right|_{\gamma=0}=\pm2\beta\partial_x p_\text{B}$, while $\left.\partial_{\boldsymbol \nu}p_\text{B}^+\right|_{\gamma=0} = \left.\partial_{\boldsymbol \nu}p_\text{B}^-\right|_{\gamma=0}$, and so the observable $p_\text{B}^+-p_\text{B}^-$ is not sensitive to the variances of $\boldsymbol \nu$, to lowest order in $\gamma$. Indeed, if we construct the Fisher matrix $F_{ab}=\sum_i \frac{N_i}{p_i(1-p_i)}\frac{\partial p_i}{\partial\vartheta_a}\frac{\partial p_i}{\partial\vartheta_b}$, then the marginalized Fisher matrix for the signal is 

\begin{equation}
    F_\gamma^\text{marg} = F_{\gamma\gamma}-F_{\gamma\boldsymbol\nu}F_{\boldsymbol{\nu}\boldsymbol{\nu}}^{-1}F_{\boldsymbol{\nu}\gamma},
\end{equation}
and to lowest order in $\gamma$ we find

\begin{equation}
    F_\gamma^\text{marg} = \frac{4\beta^2 N}{p_\text{B}(1-p_\text{B})}\pare{\frac{\partial p_\text{B}}{\partial x}}^2,
\end{equation}
where we took $N_\text{B}^+=N_\text{B}^-\approx N/2$. As we see, the red sideband atoms $N_\text{R}$ do not contribute to the local Fisher information for $\gamma$. They are instead used to calibrate $\theta$ and $\bar{n}_x$, and thereby the slope $\partial_x p_\text{B}(x,\bar{n}_x,\theta)$, potentially allowing for $N_\text{R}\ll N$, provided the required calibration precision is achieved. The ratio of the marginalized Fisher matrix under the HST protocol to the QFI is

\begin{equation}
\frac{F_{\gamma}^\text{marg}}{F_{Q,\gamma}} = \frac{\beta^2(2\bar{n}_x+1)}{p_\text{B}(1-p_\text{B})}\parea{\frac{\partial p_\text{B}}{\partial x}}^2,
\label{eq:HST_error}
\end{equation}
which we plot in Fig.~\ref{fig:FisherRatio}. In order to suppress artificially large Raman pulses, in the plot we add Gaussian noise to the area of the pulse, $\theta\to\theta(1+\epsilon)$, so that $\sin^2\pare{\frac{\theta\sqrt{n+1}}{2}}\to \frac{1}{2}\parea{1-e^{-\sigma_\epsilon^2\theta^2(n+1)/2}\cos(\theta\sqrt{n+1})}$. The error $\sigma_\epsilon$ can be interpreted as an RMS fractional Rabi-frequency drift. 

This discussion assumes a global $\theta$ and $\bar{n}_x$. Static inhomogeneities can be incorporated by averaging the site-dependent slopes $\partial_x p_\text{B}^a$. As long as the atoms are assigned randomly to each readout population per shot, and they sample the same $\boldsymbol{\nu}$ distributions, there is no systematic bias, and the additional variance scales as $\propto N^{-1}$, assuming all $+/-$ pairs are independent. Correlations and drifts can modify this scaling or introduce a systematic bias. Such effects should, ultimately, be measured, to determine the efficacy of this protocol.

%%%%%%%%%%%%
\section{Kaluza-Klein mode summation}
\label{app:extradims}

For an extra-dimensional manifold $\mathcal{M}$ whose Laplacian has discrete eigenvalues $\Lambda_\mathbf{n}$, $\Delta u = -\Lambda_\mathbf{n}u$, we can define the function $N(\Lambda)$ that counts the number of modes with eigenvalues $\leq\Lambda$, $N(\Lambda) = \sum_{\mathbf{n}}\Theta(\Lambda-\Lambda_\mathbf{n})$, 
where $\Theta$ is the Heaviside function. In this notation the KK masses are $\mu_\mathbf{n}=\Lambda_\mathbf{n}^{1/2}$. At high energies, $\Lambda\to\infty$, Weyl's law~\cite{Berard1986} gives the universal scaling 

\begin{equation}
N(\Lambda)\simeq \frac{\omega_p}{(2\pi)^p}\text{Vol}(\mathcal{M})\Lambda^{p/2},
\end{equation}
where $\omega_p$ is the volume of the unit $p$-ball in $p$ dimensions. In the presence of extra dimensions, the potential sourced by the periodic mass distribution is

\begin{equation}
    \Phi_p = 8 c_\text{d} G\Delta \rho \cos\pare{\omega_x t}\sin\pare{ q_0x}\sum_\mathbf{n} c_\mathbf{n}\frac{e^{-d\sqrt{q_0^2+\mu_\mathbf{n}^2}}}{q_0^2+\mu_\mathbf{n}^2},
    \label{eq:phiextradim}
\end{equation}
where $c_0=1$ and $c_{\mathbf{n}\neq 0}=4/3$
Defining $S(q_0,d) = \frac{1}{\text{Vol}(\mathcal{M})}\sum_{\mathbf{n}\neq 0}  \frac{e^{-d\sqrt{q_0^2+\mu_\mathbf{n}^2}}}{q_0^2+\mu_\mathbf{n}^2}$, the amplitude in Eq.~\eqref{eq:amplitude_extradim} is proportional to

\begin{equation}
\begin{split}
    S(q_0,d) 
    &\approx \frac{p\omega_p}{(2\pi)^p}\int_0^\infty \di \mu\, \mu^{p-1} \frac{e^{-d\sqrt{q_0^2+\mu^2}}}{q_0^2+\mu^2}\\
    &=\left\{\begin{matrix}
        \frac{1}{\pi q_0}\text{Ki}_1(q_0 d),\quad &p=1\\
        \frac{1}{2\pi}\text{E}_1(q_0 d),\quad &p=2\\
        \frac{q_0}{2\pi^2}(K_1(q_0 d) - \text{Ki}_1(q_0 d)),\quad &p=3
    \end{matrix}\right.
\end{split}
\end{equation}
where $\text{E}_1(z)\equiv\int_z^\infty \di t\, \frac{e^{-t}}{t}$, and $\text{Ki}_1(z)\equiv\int_z^\infty \di t \,K_0(t)$, with $K_i$ denoting modified Bessel functions of the second kind. We have ignored a $3/4$ normalization of the Newtonian contribution here, since it's already very sub-dominant. Weyl's law is everywhere appropriate as long as $q_0\mu_*^{-1}\gg 1$ as well. 

Once the sensitivity is enough to approach this threshold, the universal $N(\Lambda)\propto \text{Vol}(\mathcal{M})$ scaling no longer holds, and the result depends on the curvature of the manifold and the degeneracy of the lowest-lying modes~\cite{Kehagias_2000,Floratos_1999}. We use the exact KK mode sum in Table~\ref{tab:phases}, which differs at the tens-of-percent level compared to the Weyl law result above for Phase III, assuming a square torus $T^p$ with volume $\text{Vol}(\mathcal{M})=(2\pi R)^p$ and $\mu_\mathbf{n}^2=\sum_{i=1}^p \frac{n_i^2}{R}$, and we leave other compactifications to future work.

%%%%%%%%%%%%%%%%%%%%%%%%%%%%%
\section{Lindblad equation}
\label{app:lindblad}

For a real, stationary, zero-mean noise process $\xi(t)$, we define $C_\xi(\tau)=\avg{\xi(t+\tau)\xi(t)}$, and
\begin{equation}
    \widetilde S_\xi(\omega)
    =\int_{-\infty}^{\infty}\di\tau\,
    e^{i\omega\tau}C_\xi(\tau),
\end{equation}
the two-sided angular-frequency power spectrum. In the main text, we use the one-sided convention
\begin{equation}
    S_\xi(\omega)=
    \begin{cases}
        2\widetilde S_\xi(\omega),&\omega>0,\\
        \widetilde S_\xi(0),&\omega=0.
    \end{cases}
\end{equation}
With this definition spectral densities satisfy $S_\xi(\omega)\frac{\di\omega}{2\pi}=S_\xi(\nu)\di\nu$,
and so $S(\omega=2\pi\nu)=S(\nu)$.

Using the forcing and detuning definitions in the text, and separating a mean detuning $\bar{\delta}$ from $\avg{\Delta}=0$, the secular Born--Markov equation in the interaction picture is
\begin{equation}
    \begin{split}
        \frac{\di\rho}{\di t}
    &=-i\parea{\bar{\delta}a^\dagger a,\rho}
    +\Gamma_\phi\mathcal{D}[a^\dagger a]\rho \nonumber\\
    &\quad+\Gamma_F\pare{\mathcal{D}[a]+\mathcal{D}[a^\dagger]}\rho
    +\Gamma_{\mathrm \Delta}
    \pare{\mathcal{D}[a^2]+\mathcal{D}[a^{\dagger 2}]}\rho ,
     \label{eq:noise_lindblad}
    \end{split}
\end{equation}
where the Lindbladian is $\mathcal{D}[L]\rho
    =L\rho L^\dagger-\frac{1}{2}\pareb{L^\dagger L,\rho}$,
and
\begin{equation}
    \Gamma_F=\frac{S_F(\omega_x)}{4m\omega_x},
    \quad
    \Gamma_\phi=S_\Delta(0),
    \quad
    \Gamma_{\Delta}=\frac{S_\Delta(2\omega_x)}{8},
    \label{eq:noise_rates}
\end{equation}
where the dephasing rate is derived assuming a white $S_\Delta$ at low frequencies.

Taking the expectation value of $n=a^\dagger a$ gives Eq.~\eqref{eq:heating} of the main text. 

Finally, since
$\mathcal{D}^\dagger[a^\dagger a]a=-a/2$, the zero-frequency noise causes the coherent amplitude to decay as
\begin{equation}
    \abs{\avg{a(t)}}=
    \abs{\avg{a(0)}}\exp\parea{-\frac{1}{2}\Gamma_\phi t}.
\label{eq:coh_dephase}
\end{equation}
Including the static detuning $\bar\delta$ and assuming that $S_\Delta(\omega)$ is white for order-of-magnitude estimates, this gives Eq.~\eqref{eq:dephasing} of the main text. Keeping the a general detuning power spectrum modifies the exponent of Eq.~\eqref{eq:coh_dephase} as $\Gamma_\phi t\to \int_{-\infty}^{\infty}
\frac{d\omega}{2\pi}\,
S_\Delta(\omega)
\frac{4\sin^2(\omega t/2)}{\omega^2}$.

\bibliography{main}

\end{document}